\documentclass[12pt]{article}
\usepackage{amssymb,amsmath}
\usepackage{cite}
\numberwithin{equation}{section}

\newcommand{{\cg}}{\mbox{$\cal{G}$}}

\newcommand{\cS}{\mbox{$\cal{S}$}}
\newcommand{\cR}{\mbox{$\cal{R}$}}
\newcommand{\cT}{\mbox{$\cal{T}$}}

\newcommand{\CC}{\mbox{${\mathbb C}$}}

\newcommand{\II}{\mbox{${\mathbb I}$}}

\begin{document}
\renewcommand{\thefootnote}{\fnsymbol{footnote}}
\newpage
\pagestyle{empty} \setcounter{page}{0}

\markright{\today\dotfill DRAFT\dotfill }


\newcommand{\LAP}{LAPTH}
\def\logo{{\bf {\huge LAPTH}}}

\centerline{\logo}

\vspace {.3cm}

\centerline{{\bf{\it\Large Laboratoire d'Annecy-le-Vieux de
Physique Th\'eorique}}}

\centerline{\rule{12cm}{.42mm}}

\vfill\vfill

\begin{center}

{\LARGE  {\sffamily Factorization in integrable systems with
impurity}}

\vfill

{\large V. Caudrelier\footnote{caudreli@lapp.in2p3.fr}
\\[.21cm]
 Laboratoire de Physique Th{\'e}orique \LAP\footnote{UMR 5108
  du CNRS, associ{\'e}e {\`a} l'Universit{\'e} de
Savoie.}\\
9, chemin de Bellevue, BP 110, \\
F-74941  Annecy-le-Vieux Cedex, France.\\[.21cm]}
\end{center}

\vfill\vfill\vfill

\begin{abstract}
This article is based on recent works done in collaboration with
M. Mintchev, {\'E}. Ragoucy and P. Sorba. It aims at presenting
the latest developments in the subject of factorization for
integrable field theories with a reflecting and transmitting
impurity.
\end{abstract}

\vfill \centerline{PACS numbers: 11.10.Kk, 11.55.Ds} \vfill
\rightline{\texttt{hep-th/0508157}} \rightline{\LAP-conf-1112/05}

\newpage
\pagestyle{plain}
\setcounter{footnote}{0}

\section*{Introduction}

\hspace{0.4cm} One particular aspect of quantum integrable field
theories is the existence, by definition, of countably many
independent conserved quantities. In $1+1$ dimensions, nontrivial
massive quantum field theories can be studied non-perturbatively
thanks to the following results \cite{Parke}: there is no particle
production, the momenta are conserved individually and finally,
any $N$-particle process can be decomposed as a sequence of
two-particle processes. This last property is the so-called
\textit{factorization property}. It follows that the central
ingredient for such theories is the two-body scattering matrix
which has to satisfy the celebrated (quantum) Yang-Baxter equation
\cite{Yang,Baxter}.

It is the purpose of this article to present in a comprehensive
way how these well-known facts can be generalized to account for
the presence of a reflecting and transmitting defect, called
impurity. To this end, we will briefly recast what is known of
factorization for integrable field theories on the whole line and
on the half-line. When including an impurity, this has been
generalized in two different ways, resulting in an apparent
difficulty. We will show how to reconcile the two points of view.

\section{Factorization on the whole line}

\subsection{Physical data and Yang-Baxter equation}

\hspace{0.4cm} Let us consider a quantum integrable field theory
for massive particles with $n$ internal degrees of freedom. One of
the essential ingredients is the two-body scattering matrix whose
coefficients $\cS_{\alpha_1\alpha_2}^{\beta_1\beta_2}(k_1,k_2)$,
$\alpha_1,\ldots,\beta_2=1,\ldots,n$ are functions of the
rapidities (or momenta) $k_1$, $k_2\in\CC$ parametrizing the
dispersion relation of the two particles. This matrix encodes the
interaction between particles. For convenience, in the rest of the
article, we will use auxiliary spaces and consider the two-body
scattering matrix $\cS_{12}(k_1,k_2)$ as an element of
$End(\CC^n\otimes \CC^n)(k_1,k_2)$
\begin{eqnarray}
\cS_{12}(k_1,k_2)=\cS_{\alpha_1\alpha_2}^{\beta_1\beta_2}(k_1,k_2)~
E_{\alpha_1\beta_1}\otimes E_{\alpha_2\beta_2}\,,
\end{eqnarray}
where $E_{\alpha\beta}$, $\alpha,\beta=1,\ldots,n$ is the
canonical basis of $\CC^n$ and summation over repeated indices is
implied.

The physical unitarity of the total scattering matrix is
guaranteed by that of the two-body scattering matrix
\begin{eqnarray}
\cS_{12}(k_1,k_2)\cS^\dagger_{12}(k_1,k_2)=\II\otimes \II\,,
\end{eqnarray}
where the dagger stands for Hermitian conjugation and $\II$ is the
$n\times n$ unit matrix. In turn, this is implied by two
conditions that are usually imposed for convenience
\begin{eqnarray}
\label{unitarity_S}
\mbox{Unitarity}\qquad\qquad\qquad\quad~& & \cS_{12}(k_1,k_2)\cS_{21}(k_2,k_1)=\II\otimes \II\,,\\
\label{herm_S} \mbox{Hermitian analyticity}\qquad &
&\cS^\dagger_{12}(k_1,k_2)=\cS_{21}(k_2,k_1)\,.
\end{eqnarray}

In this context, the factorization property of the underlying
theory is represented by the Yang-Baxter equation that we require
for the two-body scattering matrix
\begin{eqnarray}
\label{YBE} \cS_{12}(k_1,k_2)\cS_{13}(k_1,k_3)\cS_{23}(k_2,k_3)=
\cS_{23}(k_2,k_3)\cS_{13}(k_1,k_3)\cS_{12}(k_1,k_2)\,.
\end{eqnarray}
This has to be understood as an identity in $End(\CC^n\otimes
\CC^n\otimes \CC^n)(k_1,k_2,k_3)$.

\subsection{Algebraic setup}

\hspace{0.4cm} It has been realized in \cite{ZZ,Fad} that the
previous two-body scattering matrix could be taken as the central
piece of an algebraic setup analogous to the Heisenberg algebra
but aiming at describing the asymptotic states of an
\textit{interacting} theory characterized by $\cS_{12}(k_1,k_2)$.
The consistency of this approach is ensured by the above
conditions imposed on $\cS_{12}(k_1,k_2)$. This is known as the
Zamolodchikov-Faddeev (ZF) algebra \cite{ZZ,Fad}. This approach
has been rigorously investigated in \cite{LM} where the Fock space
representation of the ZF algebra was explicitly constructed. The
ZF algebra has proved extremely fruitful in the Quantum Inverse
Scattering method (see e.g. \cite{Fad}).

In this context, the quantum nonlinear Schr\"odinger equation is
the paradigm of quantum integrable field theories (see e.g.
\cite{Dav} and references therein). Indeed, using the ZF algebra
for this system, it is possible to reconstruct the canonical
quantum field satisfying the nonlinear evolution equation. One
also easily gets the $N$-particle scattering matrix elements and
the time-dependent correlation functions (at zero temperature).
Finally, this algebraic approach enables one to identify the
symmetry algebra of the system \cite{MW,MRSZ}. This powerful
approach makes it very tempting to generalize the algebraic setup
when investigating systems with boundary or impurity.

\section{Generalizing factorization}

\hspace{0.4cm} Consider a quantum integrable system with a (fixed)
boundary. We are interested in the factorization property of the
theory knowing that, in addition to the two-body scattering matrix
$\cS_{12}(k_1,k_2)$, one has to take into account the reflection
matrix $\cR(k)$. The latter represents the scattering properties
of a particle characterized by $k$ with the boundary. Generalizing
the ideas of \cite{ZZ} to a system with boundary, I. Cherednik
discovered in \cite{Che} that the reflection matrix should satisfy
the \textit{reflection equation}. The factorization property means
here that the $N$-particle scattering matrix is built out of
$\cS_{12}(k_1,k_2)$ and $\cR(k)$ only. In analogy with the
properties listed in the first section for $\cS_{12}(k_1,k_2)$ one
requires the following properties for the physical data $\cR(k)$
\begin{eqnarray}
\label{unitarity_R}
\mbox{Unitarity}\qquad\qquad\qquad\quad~& & \cR(k)\cR(-k)=\II\,,\\
\label{herm_R}
\mbox{Hermitian analyticity}\qquad & &\cR^\dagger(k)=\cR(-k)\,,\\
\label{RE}\mbox{Reflection equation}\qquad~~~ &
&\cS_{12}(k_1,k_2)\cR_1(k_1)\cS_{21}(k_2,-k_1)\cR_2(k_2)\\
\qquad\qquad\qquad\qquad & &=\cR_2(k_2)
\cS_{12}(k_1,-k_2)\cR_1(k_1)\cS_{21}(-k_2,-k_1)\nonumber\,.
\end{eqnarray}

\subsection{Algebraic point of view}

\hspace{0.4cm} As explained above, it is very promising to design
an algebraic approach for systems with boundary. Two approaches
have been developed so far. The first one is based on a
\textit{boundary operator} \cite{GZ,FK} which is added to the
usual ZF generators. The second and more recent approach relies on
the so-called \textit{boundary algebra} \cite{LMZ} which also
includes an additional generator accounting for the boundary but
deeply modifies the defining relations of the original ZF algebra.
Each approach has its own interests. For example, the first one
has been convenient for the bootstrap program with boundary (see
e.g. \cite{GZ,CDRS}) while the second is most useful in the
quantum inverse scattering method with boundary (see e.g.
\cite{GLM}). Let us stress however that there is no rigorous
justification of the first approach (as admitted by the authors of
\cite{GZ} themselves) as opposed to the second one \cite{LMZ}.

Nevertheless, there is no difference between the two approaches at
the level of the equations imposed on the physical data. Both
consistently reproduce the sets of relations
(\ref{unitarity_S}-\ref{YBE}) and (\ref{unitarity_R}-\ref{RE}).

Both approaches were naturally generalized to the case of a defect
(or impurity). In this case, the possibility for transmission is
encoded in an additional matrix: the transmission matrix $\cT(k)$.
At the algebraic level, the differences are even more important
than in the boundary case. Indeed, in the first approach, the
boundary operator is replaced by a \textit{single defect operator}
added to the usual ZF algebra, giving rise to a \textit{defect
algebra} \cite{DMS} while in the second approach, the boundary
algebra is turned into a \textit{reflection-transmission algebra}
where \textit{two} additional generators account for the presence
of the impurity \cite{MRS}. The status of the two approaches is
parallel to the boundary case. The first one remains quite formal
and has been used for various computations of statistical physics
\cite{DMS}. Let us stress however that a strong restriction
discovered in \cite{CFG} holds in this context, limiting the study
of integrable systems with reflection and transmission essentially
to noninteracting ones. The second approach stands on mathematical
foundations involving the explicit construction of Fock
representations \cite{MRS}. It proved fundamental in the quantum
inverse scattering method applied to the nonlinear Schr\"odinger
equation with impurity \cite{CMR}, which constitutes the first
known example of this kind, and in the investigation of the
corresponding symmetry algebra \cite{CR}.

Nevertheless, the situation is completely different from the
boundary case as soon as one is interested in the factorization
property for the physical data since the two approaches seem to
give different physical equations. This is explained in the rest
of this article.

\subsection{Going back to physical data}

\hspace{0.4cm} We are concerned with the analogs of the
Yang-Baxter and reflection equations when transmission is allowed.
Let us emphasize that these \textit{reflection-transmission
quantum Yang-Baxter equations} (RTQYBE) are the crucial elements
for physics in the sense that they encode the factorization
property of any $N$-particle process and allow for physical
computations. The previous algebraic setups are convenient
theoretical tools for which the only constraints are
self-consistency and consistent reproduction of the RTQYBE.

Let us focus now on the equations obtained in the first approach.
To be accurate, introduce the Lorentz (Galilean) invariant
two-body scattering matrix $S_{12}(k_1-k_2)$. Introduce also two
reflection matrices $R^+(k)$, $R^-(k)$ and two transmission
matrices $T^+(k)$, $T^-(k)$, as is done in \cite{CFG}. They
account for possible different bahaviours on the left and on the
right of the impurity. Invoking the consistency and the
associativity of the corresponding defect algebra, one recovers
the fundamental relations (\ref{unitarity_S}-\ref{YBE}) and
deduces the following typical relations
\begin{eqnarray}
\label{eq_ref_DMS}
S_{12}(k_1-k_2)R^+_1(-k_1)S_{21}(k_1+k_2)R^+_2(-k_2)\qquad\qquad\qquad\qquad
\nonumber\\
=R^+_2(-k_2)
S_{12}(k_1+k_2)R^+_1(-k_1)S_{21}(k_1-k_2)\,,\\
T^+_1(k_1)S_{21}(k_2-k_1)R^+_2(-k_2)=R^+_2(-k_2)T^+_1(k_1)S_{21}(-k_2-k_1)\,,\\
S_{12}(k_1-k_2)T^+_1(k_1)T^+_2(k_2)=T^+_2(k_2)T^+_1(k_1)S_{12}(k_1-k_2)\,,\qquad\\
\label{unitarite_DMS} R^+(-k)R^+(k)+T^-(k)T^+(k)=\II~~,~~
R^-(k)T^+(k)+T^+(-k)R^+(k)=0\,.
\end{eqnarray}

We refer the reader to the original works for the remaining
relations. What is important here is to note that the relations
involving $S_{12}$ and $T^+$ are \textit{cubic}, in that they
involve three terms on each side.

Now, let us move on to the equations obtained from the second
approach. The physical data is given by $\cS_{AB}(k_1,k_2)$,
$\cR(k)$, $\cT(k)$ (where we labelled the auxiliary spaces by
letters for later convenience). One may wonder about the
possibility of different left and right reflection and
transmission but we will see in the example below that this is
intrinsically encoded in this approach. The sought relations
appear as Fock representations of the RT algebra \cite{MRS} and
read
\begin{eqnarray}
\label{eq_ref_RT}
\cS_{AB}(k_1,k_2)\cR_A(k_1)\cS_{BA}(k_2,-k_1)\cR_B(k_2)\qquad\qquad\qquad\qquad
\nonumber\\
=\cR_B(k_2) \cS_{AB}(k_1,-k_2)\cR_B(k_1)\cS_{BA}(-k_2,-k_1)\,,
\end{eqnarray}
\begin{eqnarray}
\label{RT}
\cS_{AB}(k_1,k_2)\cR_A(k_1)\cS_{BA}(k_2,-k_1)\cT_B(k_2)\qquad\qquad\qquad\qquad
\nonumber\\
=\cT_B(k_2) \cS_{AB}(k_1,k_2)\cR_A(k_1)\cS_{BA}(k_2,-k_1)\,,
\end{eqnarray}
\begin{eqnarray}
\label{TT}
\cS_{AB}(k_1,k_2)\cT_A(k_1)\cS_{BA}(k_2,k_1)\cT_B(k_2)\qquad\qquad\qquad\qquad
\nonumber\\
=\cT_B(k_2) \cS_{AB}(k_1,k_2)\cT_A(k_1)\cS_{BA}(k_2,k_1)\,,
\end{eqnarray}
\begin{eqnarray}
\label{unitarite_RT_mat}
\cR(k)\cR(-k)+\cT(k)\cT(k)=\II~~,~~\cR(k)\cT(-k)+\cT(k)\cR(k)=0\,.\qquad
\end{eqnarray}

Here the equations involving transmission are \textit{quartic}. It
seems that the two approaches are incompatible
 as long as one is interested in the physical equations for reflection and
 transmission. But how to decide which one is correct?

One can notice that both approaches agree on the reflection
equation ((\ref{eq_ref_DMS}) and (\ref{eq_ref_RT})) and on the
unitarity relations ((\ref{unitarite_DMS}) and
(\ref{unitarite_RT_mat})). Forgetting the algebraic setup, one can
think of starting from these equations to get some insight. This
has been first presented in \cite{MRS} and detailed in
\cite{CMRS}. It is argued that solving (\ref{unitarite_RT_mat})
for $\cT(k)$ in terms of $\cR(k)$ and using the reflection
equation, one gets that $\cT(k)$ must satisfy
 the quartic relations (\ref{RT}) and (\ref{TT}) and not the cubic ones.

\subsection{Reconciling points of view}

\hspace{0.4cm} The above conclusions were at the heart of an
apparent controversy. In \cite{CMRS}, the situation has been fully
clarified. It appears that there is \textit{no contradiction}
since the first approach is merely a particular case of the second
one. Indeed, making the following particular choice for the
scattering data of the second approach\footnote{Here, if the
labels $1$ and $2$ correspond to the auxiliary space $\CC^n$ then
the labels $A$ and $B$ correspond to the auxiliary space
$\CC^2\otimes \CC^n$.}
\begin{eqnarray}
\cS_{AB}(k_1,k_2)=\left(\begin{array}{cccc}
S_{12}(k_1-k_2) &0&0&0\\
0&\II\otimes \II &0&0\\
0&0&\II\otimes \II&0\\
0&0&0& S_{21}(k_2-k_1)
\end{array}  \right)\,.
\end{eqnarray}
\begin{eqnarray}
\cR(k)=\left(\begin{array}{cc} R^+(-k) & 0 \\
0 & R^-(-k)
\end{array}\right)~~,~~
\cT(k)=\left(\begin{array}{cc} 0 & T^-(k)  \\
T^+(k) & 0
\end{array}\right)\,,
\end{eqnarray}
and plugging in (\ref{eq_ref_RT}-\ref{unitarite_RT_mat}), one
recovers the equations for the data of the first approach.

\section*{Conclusions}

\hspace{0.4cm} The reflection-transmission quantum Yang-Baxter
equations have the form (\ref{eq_ref_RT}-\ref{unitarite_RT_mat}).
They are naturally obtained in the Reflection-Transmission (RT)
algebras approach \cite{MRS} and reproduce as a particular case
the original equations obtained in \cite{DMS,CFG}. However, they
do not suffer from the same limitation and allow for quantum
\textit{interacting} integrable systems with reflection
\textit{and} transmission. This was first illustrated in
\cite{CMR} using the well-known nonlinear Schr\"odinger equation.
It appears that the way out of free models is \textit{not} due to
the more general dependence in the momenta allowed for in the RT
approach. This is discussed and illustrated in \cite{CMRS} where
explicit solutions with a Lorentz invariant two-body scattering
matrix are established. The key point lies in the quartic
relations involving transmission. In this respect, we remark that
the physical data used to solve the nonlinear Schr\"odinger
equation with impurity satisfies the quartic but not the cubic
relations.

This shows that the RT algebra approach offers new interesting
possibilities in the study of integrable systems with impurity.

\bigskip
{\small The author is grateful to the organizers of the $XIV^{th}$
International Colloquium on Integrable Systems held in Prague in
June 2005.}

\end{document}